\documentclass[slac_one]{revtex4}
\usepackage{graphicx}
\usepackage{fancyhdr}
\newcommand{\kev}{\text{keV}}
\newcommand{\mev}{\text{MeV}}
\newcommand{\gev}{\text{GeV}}

\newcommand{\eqref}[1]{Eq.~(\ref{#1})}

\begin{document}

\title{{\small{LEPTON - PHOTON 2007}}\\ 
\vspace{12pt}
Dark Matter Decaying Now } 

\author{Jose~A.~R.~Cembranos}
\affiliation{Department of Physics and Astronomy, University of
California, Irvine, CA 92697, USA}

\author{Jonathan L.~Feng}
\affiliation{Department of Physics and Astronomy, University of
California, Irvine, CA 92697, USA}

\author{Louis E.~Strigari}
\affiliation{Department of Physics and
Astronomy, University of California, Irvine, CA 92697, USA}

\begin{abstract}
The instability of dark matter may produce visible signals in the 
spectrum of cosmic gamma-rays.  We consider this possibility in frameworks 
with additional spatial dimensions and supersymmetry. Examples of particles
include superweakly-interacting massive particles such as gravitinos in 
supersymmetry models, the lightest Kaluza-Klein (KK) state in models with 
universal extra dimensions, and weakly-interacting massive particles such as 
branons in flexible brane-worlds.
\end{abstract}

\maketitle


\thispagestyle{fancy}

\section*{Introduction}

Though we know that approximately one fourth of the energy density of the 
Universe is made of non standard particles, the microscopic nature of such
particles remains as one of 
the major mysteries in science. This fundamental piece of standard cosmology, 
known as dark matter (DM), is usually assumed to be in the form of stable, massive, 
collisionless and non-self-interacting particles. Weakly-interacting massive 
particles (WIMPs) that freeze-out with the right thermal abundance in the early 
Universe are well-motivated dark matter candidates with masses and interaction 
cross sections of the order of the weak scale. WIMPs emerge naturally from 
different particle physics scenarios such as the lightest supersymmetric particle 
(LSP) in R-parity conserving supersymmetry (SUSY) models~\cite{SUSY}, the lightest 
Kaluza-Klein (KK) excitation (LKP) in models with universal extra dimensions 
(UED)~\cite{UED}, or branons in flexible brane-worlds~\cite{BW1,BW2}.

Other well established candidates are superweakly-interacting 
massive particles (superWIMPs), which are born from unstable WIMPs with typical 
lifetimes $\tau \sim 1$ s - $10^8$ s, and naturally inherit the abundance of the 
WIMPs from which they are produced \cite{Feng:2003xh,Feng:2003uy}. 
Examples of superWIMPs include non-thermally produced weak-scale
gravitinos~\cite{Feng:2003xh,Feng:2003uy,Ellis:2003dn,Feng:2004zu,%
Roszkowski:2004jd}, axinos~\cite{axinos}, and
quintessinos~\cite{Bi:2003qa} in supersymmetry; and Kaluza-Klein
graviton and axion states in models with universal extra
dimensions~\cite{Feng:2003nr}.

The experimental search for DM depends on its nature, and in general, a strong 
statement would need the interplay of collider experiments \cite{CollSW, Coll} and 
astrophysical observations. These latter observations are typically classified 
as direct or indirect searches (see \cite{others} however, for different
alternatives). Additionally, elastic scattering of DM particles
from nuclei should lead directly to observable nuclear recoil signatures.

In this note we point out that present DM particles can be unstable and have associated 
lifetimes of the order of the age of the Universe or longer. 
In this case, new DM signatures can be explored, such as anomalies in cosmic gamma-rays.
Such DM particles can be found in the above mentioned scenarios with supersymmety 
and extra dimensions, not only in form of superWIMPs, but also in form of WIMPs, for
example in the case of branons.

\section*{Gamma-ray background}

For highly degenerate particles, the decays can take place very late
to soft photons.  The photon spectrum is not thermalized and produces bumps in 
the diffuse photon spectrum that may be observable.  The present
differential flux of photons from a general decay is given by
\begin{equation}
\frac{d\Phi}{ dE_{\gamma}}
=\frac{c\,n_\gamma}{4\pi} \int_0^{t_0} dt
\frac{N(t)}{V_0}
\frac{d\Gamma_{\gamma}}{d\varepsilon_\gamma}
\frac{d\varepsilon_\gamma}{dE_{\gamma}},
\end{equation}
where $n_\gamma$ is the number of photons produced in a single decay.
$t_0 \simeq 4.3\times 10^{17}$ s. is the age of the
universe, $N(t) = N^{\text{in}}
e^{-t/\tau}$, where $N^{\text{in}}$ is the initial number of
decaying particles, $V_0$ is the present volume of the
universe, and the relation between the produced energy $\varepsilon_\gamma$,
and observed energy $E_{\gamma}$, is given by the scale factor of the 
Universe: $\frac{d\varepsilon_\gamma}{dE_{\gamma}}=(1+z)\equiv a^{-1}$. 
In the case of a two body decay, the 
produced photons have a determined energy given by the 
mass splitting between the decaying particle and the produced one:
$\varepsilon_\gamma=\Delta M$. It means that 
$\frac{d\Gamma_{\gamma}}{dE_{\gamma}}=\delta \left( E_{\gamma} - \frac{\varepsilon_\gamma}{1+z} \right)\Gamma_{\gamma}$ \cite{Feng:2003xh,Cembranos:2007fj} and
\begin{equation}
\frac{d\Phi}{ dE_{\gamma}} =
\frac{c}{4\pi} \frac{N^{\text{in}}\,
e^{-P(E_\gamma /\varepsilon_{\gamma})/\tau}}
{V_0 \tau} \, \left ( \frac{1}{aH} \right )_{E_\gamma/\varepsilon_\gamma}
 \, \Theta(\varepsilon_\gamma - E_{\gamma}) \,, 
\label{eq:extragalacticphi}
\end{equation}
where 
\begin{equation}
P(a)\equiv \frac{2\left(
\ln{\left[\sqrt{\Omega_{\Lambda}a^3}
+\sqrt{\Omega_{M}+\Omega_{\Lambda}a^3}\right]}
-\ln{\left[\sqrt{\Omega_{M}}\right]}
\right)}{3 H_0 \sqrt{\Omega_{\Lambda}}}
\label{eq:P}
\end{equation}
is given by the equation:
\begin{equation}
\frac{da}{dt}= a H
\simeq H_0\sqrt{
\frac{\Omega_{M}}{a}
+\Omega_{\Lambda}a^2} 
\equiv Q(a)
\,,
\label{qveq}
\end{equation}

\begin{figure}
\resizebox{3.56in}{!}{
\includegraphics{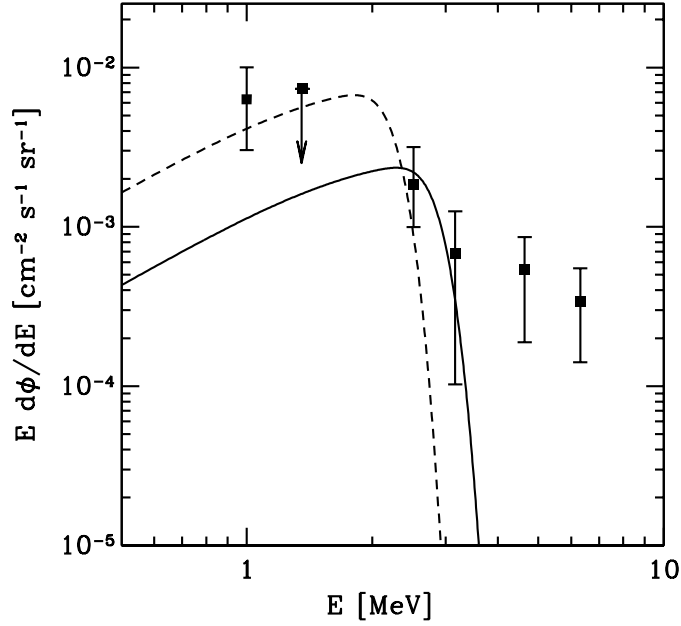}
} 
\caption{Contributions from extragalactic dark matter decays along
with data for the the cosmic gamma background observed by COMPTEL. 
The curves are for $B^1$ decay in mUED with lifetime 
$\tau= 10^3 t_0$ and $m_{B^1}= 800 \gev$ (solid) and, $\tilde B$ decay
in SUSY with lifetimes $\tau= 5 \times 10^3 t_0$ and  
$m_{\tilde B}= 80 \gev$ (dashed). We have assumed that these particles
make up all of non-baryonic dark matter ($\Omega_{NBDM}=0.2$) and smeared
all spectra with energy resolution $\Delta E/E= 10\%$, characteristic of
COMPTEL \cite{Cembranos:2007fj}.
\label{fig:cgb} }
\end{figure}
The flux has a maximum at 
\begin{equation}
E_{\gamma}= \varepsilon_{\gamma}
\left[\frac{\Omega_{M}}{2\Omega_{\Lambda}}
U(H_0^2\tau^2\Omega_{\Lambda})\right]^{\frac{1}{3}}\,,
\label{eq:fluxmax}
\end{equation}
where
$U(x)\equiv \frac{x+1-\sqrt{3x+1}}{x-1}$.
Here we are neglecting the radiation content,
$\Omega_{R}\sim 0$, and the curvature: $k\sim 0$ \cite{Cembranos:2007fj}. For  this calculations, 
we use a flat cosmological model with $\Omega_{m} = 0.3$, $\Omega_\Lambda = 0.7$, 
and $h =0.7$. An interesting side point of this analysis is the sensitivity of the
signal to cosmology. The gamma ray background may provide novel 
constraints on the above parameters and dark energy properties.

We can gain further insight into the spectrum shape in Fig. 
\ref{fig:cgb} by considering asymptotic limits. 
If the lifetime is much shorter than the age of the Universe,
$H_0^2\tau^2\Omega_{\Lambda}<<1$, the flux grows as 
$d\Phi/dE_{\gamma} \propto E^{1/2}$ until it reaches its maximal value at:
\begin{equation}
E^{\text{max}}_{\gamma}\simeq \varepsilon_{\gamma}
\left[\frac{\Omega_{M}H_0^2\tau^2}{4}
\left(1-5\frac{H_0^2\tau^2\Omega_{\Lambda}}{4}
+...\right)\right]^{\frac{1}{3}}\,.
\label{max}
\end{equation}
From this energy, the flux is suppressed exponentially due to the decreasing
number of decaying particles \cite{Feng:2003xh}. On the other hand, if the lifetime is much
longer than the age of the Universe, $H_0^2\tau^2\Omega_{\Lambda}>>1$, 
the flux only grows as  $d\Phi/dE_{\gamma} \propto E^{1/2}$ when the dark matter
decays in the matter dominated epoch. For decays in the vacuum dominated Universe, 
the flux decreases as $d\Phi/dE_{\gamma} \propto E^{-1}$, reaching its maximum at the 
transition between these two regimes:
\begin{equation}
E^{\text{max}}_{\gamma}\simeq \varepsilon_{\gamma}
\left[\frac{\Omega_{M}}{2\Omega_{\Lambda}}
\left(1-\frac{\sqrt{3}}{H_0\tau\sqrt{\Omega_{\Lambda}}}
+...\right)\right]^{\frac{1}{3}}\,.
\label{eq:vacuummax}
\end{equation}
Thus at first order the maximum does not depend on $\tau$ \cite{Cembranos:2007fj}. 

\begin{figure}
\resizebox{3.56in}{!}{
\includegraphics{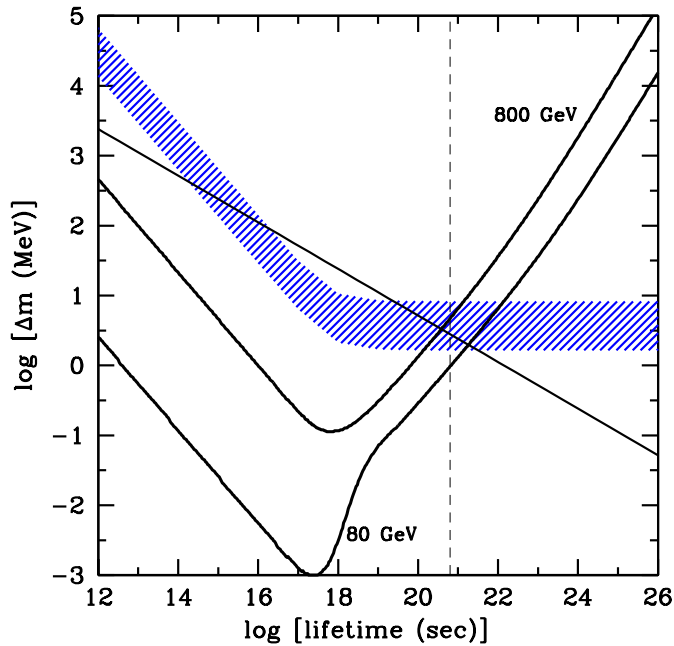}
} 
\caption{The general mass difference-lifetime parameter space for 
decaying dark matter scenarios. The solid lines reflect the points where
the peak of the spectrum from decays exactly matches the observed 
photon spectrum. The shaded region corresponds to a detected energy
in the unaccounted for range 1-10 MeV. The diagonal line shows the 
relation for the $G^1$ LKP model in mUED \cite{Cembranos:2007fj}. 
\label{fig:parameterspace} }
\end{figure}

In figure~\ref{fig:parameterspace}, we show the generic region of parameter space
$\Delta m$-$\tau$ that is probed by the diffuse photon background. We use the
limit on the diffuse photon background from $\sim$ keV-100 GeV as determined
in \cite{Gruber:1999yr}. The solid lines label the points where the peak of the photon 
spectrum from decays exactly matches the observed photon background. Of course,
for energy regimes where the photon background is well-resolved, these lines 
provided a conservative upper limit to the amount of photon flux allowed from 
dark matter decays. For a
fixed decaying DM (DDM) mass scale, all points above these lines are roughly inconsistent with 
the photon background, as the peak of the decay photon spectrum lies above the
observational limits. For lifetimes $\tau \gtrsim t_0$, the peak of the energy spectrum
is very similar to the mass difference. In this limit the relationship between the 
observed peak and the mass difference 
is given by Eq.~(\ref{eq:vacuummax}). The turnover for lifetimes 
$\lesssim t_0$ reflects the fact that the photons of an emitted energy redshift to appear
at different observed energies. For example, the decay of an 80 GeV particle with
lifetime $\tau \simeq 10^{12}$ sec. ($z \simeq 4000$) and $\Delta m \simeq$ MeV
produces a spectrum with peak $\sim$ keV. 

The shaded band in figure~\ref{fig:parameterspace} shows the region where the 
energy of the detected photons fall in the unaccounted for range of 1-10 MeV. 
The diagonal line shows the relation for the model $B^1 \rightarrow G^1 + \gamma$, 
though similar relations hold for the other gravitational suppressed two-body decays. Importantly, these relations are independent of the mass scale for these high degeneracies.  
The region of parameter space defined by $\tau \simeq 10^{21}$ sec and $\Delta m 
\simeq$ MeV defines the preferred mUED and SUSY models that can account for
the cosmic gamma-ray background in the 1-10 MeV regime. 
 
\section*{Gamma-ray lines from the center of the Galaxy} 

If the DDM comprises the majority of the dark matter today, they will decay inside our local halo producing photons, positrons, electrons and neutrinos. The fluxes of all these particles will reach the Earth
providing the possibility of a directional dependent DM signal. If this DDM produce photons directly from two-body decays, these photons will appear as a line flux at energy exactly determined by the mass splitting, with no redshifting. To determine the Galactic signal, we define $\Psi$ as the angle between the direction of the Galactic center and the line of observation, and $l$ as the distance from the Sun to any point in 
the halo.  In a line-of-sight direction determined by $\Psi$, 
the differential $\gamma$-ray flux from decays of dark matter is 
\begin{equation}
\frac{d \Phi}{d \Omega dE_\gamma} = \frac{dN_\gamma}{dE_\gamma} \cdot 
e^{-t_0/\tau} \cdot \frac{1}{4 \pi \tau m_x} \int_0^{l_{max}(\Psi)} \rho [r(l)] dl(\Psi).   
\label{eq:totalflux}
\end{equation}
The radial distance $r$ is measured from the Galactic center, and is related to $l$
by $r^2 = l^2 + D_\odot^2 -2D_\odot l \cos \Psi$, where $D_\odot = 8.5$ kpc is
the distance from the Sun to the center of the Galaxy. The distance from the Sun
to any the edge of the halo in the direction $\theta$ is 
$l_{max} = D_\odot \cos \theta + \sqrt{r^2-D_\odot^2 \sin \theta}$. 
Each decay produces one photon with energy given by the mass difference, $\Delta m$, 
so the photon spectrum is $dN_\gamma/dE_\gamma = \delta(E_\gamma-\Delta m)$. 

The photon flux is maximized in the direction of the Galactic center, and 
must be averaged over the solid angle of the detector. 
For detectors with sensitivities to energies in the tens of MeV regime, the
solid angles are typically of order $\Delta \Omega = 2 \pi ( 1 - \cos \Psi ) \simeq 10^{-3}$. 
One can model the dark halo of the Milky Way with the results of \cite{Klypin:2001xu}. We have found that the effects of dark matter clumps in the halo are negligible. Indeed, for realistic distributions of the clumps, the flux is only increased by $\sim 1\%$ for all lifetimes. 

Positrons may also be produced in the decays. If positrons are produced with energies
smaller than $~3 \mev$, they will be effectively stopped before annihilating with electrons, 
losing their kinetic energy through collisional ionization or excitation in neutral Hydrogen 
and by interaction with plasma waves in ionized interstellar medium. In such a case, they 
produce a narrow photon line at $\epsilon_\gamma\simeq m_e\simeq 511 \kev$ if they 
annihilate directly into 2 photons or through parapositronium formation.
Annihilation in such a state happens $25\%$ of the time that positronium is formed. The other 
$75\%$ of the time, the annihilation takes place in the orthopositronium state, yielding a photon 
continuum with $\epsilon_\gamma<511$ keV. Consequently, the total number of 511 keV 
photons produced per unit time is given by 
$d n_{511}/dt= 2(1-3p/4)n_{DDM} \Gamma_{e^+}= 2(1-3p/4)\rho_{DDM} \Gamma_{e^+}/M$, 
where we are supposing that annihilation takes place through positronium formation a fraction
$p$ of the time.  

The predicted distribution of the 511 line for any particular model of the
Milky Way dark halo can be computed as the integral along the
line of sight, as a function of galactic longitude $l$ and latitude
$b$, of the emissivity $d n_{511}/dt$:
\begin{equation}
\frac{d\Phi_{511}(b,l)}{dE_{\gamma}d\Omega} \simeq
\frac{1}{4 \pi}\int_0^\infty \frac{d n_{511}(r(s,b,l))}{dt} ds\,,
\label{511}
\end{equation}
where in this case, we are parameterizing the halo radius as: 
$r(s,b,l)= D_\odot^2 + s^2 + 2D_\odot s\cos{b}\cos{l}$.
The morphology of the emission depends on the Milky Way halo profile
(in addition to the precise distribution of baryons and
their chemical composition, which determine the positron propagation). 
Observations are just sensitive to the inner region of the halo, and demand 
a cuspy halo. The intensity of the emission is so low in outermost regions 
that it is difficult to discriminate from the instrumental background.
Interestingly, the splittings and lifetimes that can account for the cosmic gamma-ray 
background at MeV energies are also appropriate to explain the 511 line. An example
of a DM candidate that can provide such phenomenology is the branon, due to its
universal coupling with all standard model particles \cite{progress}. 

\section*{Conclusions}

The main idea of this note is that decaying dark matter can provide indirect 
detection signals at scales much smaller than their masses. We have analyzed 
this idea for the extragalactic photon background observed by COMPTEL 
\cite{Weidenspointner} and the photon lines coming from the Galactic 
center observed by INTEGRAL \cite{Jean:2003ci}. 
Below $\sim 1$ MeV and above $\sim 10$ MeV, the extragalactic photon
fluxes are well-modeled by the diffuse cosmological emission of 
unresolved Active Galactic Nuclei \cite{Pavlidou:2002va}, but 
in the entire regime of  $\sim 1-5$ MeV, no population of known 
astrophysical sources are able to account  for the observed cosmic gamma-ray
background \cite{Strigari:2005hu,Ahn:2005ws}. An additional intriguing 
 source of non-relativistic positrons from the center of our Galaxy 
 has been determined recently  by the INTEGRAL satellite.

We have shown that DDM provides a viable explanation that deserves further
investigation. As we have seen, these late decays at MeV scales are natural 
for gravitationally suppressed superWIMPs, and also for TeV WIMPs such as 
branons in flexible brane-worlds.

\begin{acknowledgments}
The work of JARC and JLF is supported in part by NSF CAREER grant 
No.~PHY--0239817, NASA Grant No.~NNG05GG44G, and the Alfred 
P.~Sloan Foundation. The work of JARC is also supported by the FPA 
2005-02327 project (DGICYT, Spain). LES and JARC are supported in part by a Gary 
McCue Postdoctoral Fellowship through the Center for Cosmology at UC 
Irvine.
\end{acknowledgments}

\end{document}